\documentclass[onecolumn,11pt]{article}
\usepackage{txfonts}
\usepackage{graphicx}
\usepackage{dcolumn}
\title{Constraints and Soliton Solutions for the KdV Hierarchy and AKNS Hierarchy}
\author{NianHua Li$^{a}$,
YuQi Li$^{a}$ \\
\footnote{E-mail:liyuqi@nbu.edu.cn}
\small a. Center for Nonlinear Science Center, Ningbo University, Ningbo 315211, China}

\begin{document}

\maketitle

 {\bf Abstract:}  It is well-known that the finite-gap solutions of the KdV equation
 can be generated by its recursion operator.
 We generalize the result to a special form  of Lax  pair,
 from which a method to constrain the integrable  system to a
 lower-dimensional or fewer variable integrable system is proposed.
 A direct result is that the $n$-soliton solutions of 
the KdV hierarchy can be completely depicted
 by a series of ordinary differential equations (ODEs), 
which may be gotten by a simple but unfamiliar Lax pair.
Furthermore the AKNS hierarchy is constrained to a series of
univariate integrable hierarchies. The key is a special form of Lax
pair for the  AKNS hierarchy.  
It is  proved that under the constraints all equations of the AKNS hierarchy are linearizable.
   
\section{ Introduction}

\hspace*{6mm}Much effort has been devoted to finding the exact solutions to integrable systems
since Gardner, Greene, Kruskal and Miura found the inverse scattering (IST) 
transform method to solve the KdV.
In general on the full line
the reflectionless potential is solved by the IST  as soliton or multi-soliton solutions.
It is widely believed that the IST is inefficient to obtain the other kinds of solutions.
Therefore, from this point of view it is amazing that the periodic KdV is
completely solved by the algebraic-geometric solutions.
The crucial fact in obtaining the algebraic-geometric solutions
is that  the stationary  solutions of any higher-order KdV is  invariant to the usual KdV.
The idea that solving  PDEs in its  finite-dimensional invariant subspaces somehow 
has  been developed to the method of
{\sl nonlinearization  of Lax pair}\cite{Cao} or {\sl symmetry constraint}\cite{ChengYandL}.
In most cases, the solutions obtained by nonlinearization  of Lax pair are algebraic-geometric solutions, 
from which  useful information is hard to get because  of the complex expressions of the solutions.
Furthermore, returning to get the soliton solutions
some proper constraints has to be imposed for the algebraic-geometric solutions.
This seems to be  not straightforward.
It is even more complicated to characterize the other kinds of solutions such as
the elliptic solutions of the KdV. What is even worse is that there is no a rigid 
classification for the kinds of solutions for integrable partial differential equations (IPDEs). 
In this paper we will  neatly characterize the soliton solutions by a
less popular Lax pair for the KdV hierarchy without any knowledge of the algebraic-geometric solutions
or any other old methods for obtaining the soliton solutions such as the B\"acklund or Darboux transformations.

For an IPDE the kind of solutions that can be gotten by only solving some linear PDE 
are of special interest.
At first  glance there seems to be little chance  to realize this for an  S-integrable system such as KdV or AKNS.
The most desirable  situation for solving a nonlinear partial differential equation  
is that it can be linearized  by an appropriate change of variables, which is called  C-integrable.
The famous example of such kind is the Burgers equation.
But most researchers firmly believe that  a  true S-integrable system such as KdV or AKNS
will never be linearized by  a common  change of variables.
So it will be very interesting to know to what an extent the  S-integrable system
is solvable by the change of variables. 
In  this paper we will give a series of constraints on the AKNS hierarchy.
The final result is that under the constraints the resulting equations are all linearizable
by some proper transformations.

The paper is organized  as follows.
Section 2 introduces a special form of Lax pair, about which a basic theorem is given. 
The theorem states how to construct invariant manifolds corresponding to the special form of Lax pair.
Section 3 introduces a special form of Lax pair generating the KdV hierarchy.
By the theorem introduced in Section 2, 
the  invariant manifolds corresponding to the Lax pair are constructed.
It turns out that the invariant manifolds  are just the $n$-soliton solutions of the KdV hierarchy.
Section 4 deals with a special form of Lax pair of the AKNS hierarchy. 
We will first obtain the special kind of constraints.
Then we will solve the first few invariant manifolds in detail.
At last  we will prove the main theorem for the constraints of the AKNS hierarchy.

\section{Constraint for evolution equations with a special form of Lax pair}
\hspace*{0.6cm}Integrable equations are consistency conditions of the Lax pair
\begin{eqnarray}
  &&\hat{L}\varphi =\lambda \varphi ,\label{laxL} \\
   &&\varphi_{t}=\hat{P}\varphi \label{LaxP},
\end{eqnarray}
where the  eigenfunction $\varphi=\varphi(x,t)$ is 
$n$-dimensional vector and linear operators $\hat L$ and $\hat P$
are differential polynomials of potential $u=u(x,t)$.
The method of nonlinearization of Lax pair or symmetry constraint method set up additional constraints
between the potential $u$ and the eigenfunction $\varphi$. 
With the additional constraints Equations (\ref{laxL}) and (\ref{LaxP}) will become ODEs in most cases.
Let us still take the KdV as an example. 
The KdV $u_t=6 u u_x+ u_{xxx}$ has a well-known Lax pair 
\begin{eqnarray}
(\partial^2+u) \varphi_i&=&\lambda_i \varphi_i,  \label{KdVL} \\
\partial_t \varphi_i&=&(4 \partial^3+3 (u \partial+\partial u))\varphi_i \label{KdVP}
\end{eqnarray} 
and the well-known constraint for the KdV
is \begin{eqnarray}
u=c_0+\sum_{i=1}^n c_i \varphi_i^2. \label{KdVConstranit}
\end{eqnarray}
With the constraint (\ref{KdVConstranit}),  Equation (\ref{KdVL}) and 
Equation(\ref{KdVP}) become two sets of ODEs.
In most cases the  efficient way to find a constraint for  Lax equations
is the symmetry constraint method. 
But symmetry constraints are not all constraints.
It is observed in \cite{LLL} that systems with 
the following special form of Lax pair have  natural constraints.
\begin{itemize}
\item Operator $\hat L$ has form 
$\hat L=\hat L_++\sum_{i=1}^n h_i \partial^{-1}g_i$,
where $\hat L_+$ is a differential operator and $h_i$ and $g_i$  are
differential polynomials of potential $u$.
\item Operator $\hat P$ is a differential operator.
\end{itemize}
The following theorem guarantees a natural constraint.

{\bf Theorem 2.1} For systems with Lax pairs in the above form,
there is a  function $ L_{F} =
\sum_{i=1}^{n}a_{i}h_{i}$  such that 
a  constraint   $L_{F}=\sum_{j=1}^m b_{j}\varphi_j$ exists,
where $\varphi_j$ is the eigenfunction $\hat{L}\varphi_j=\lambda_j \varphi_j$,
$a_i$s are some proper constants, $b_j$s are arbitrary constants and $m$ is
an arbitrary positive integer.

It is also well-known that Lax pair has a less popular form
\begin{eqnarray}
&& f_{j+1}  = \hat{L}f_{j}, \label{laxfL}\\
&& f_{jt} = \hat{P}f_{j}. \label{laxfP}
\end{eqnarray}
Correspondingly Theorem 2.1 has a variant form:

{\bf Theorem 2.2} There is a  function $ L_{F} =
\sum_{i=1}^{n}a_{i}h_{i}$  such that
a  constraint   $\sum_{j=1}^m b_{j} f_j=0$ exists,
where $f_1=L_F$ and the requirements for
$\hat L$, $\hat P$, $a_i$s, $b_j$s are the same as in  Theorem 2.1.

In the following paper Theorem 2.2 will be applied more frequently.

\section{The soliton constraint for the  KdV hierarchy}
\hspace*{0.6cm}The KdV hierarchy is defined by its recursion operator
$\hat{\varphi}_{1}=\partial^2+4u+2u_{x}\partial^{-1}$ 
\begin{equation}
u_t=\hat{\varphi}_{1}^{n}u_{x},  \label{KdVhch}
\end{equation}
where $n$ is an arbitrary positive integer.  The first nontrivial equation of
the hierarchy is the KdV equation
\begin{equation}
u_{t}=6uu_{x}+u_{xxx}. \nonumber
\end{equation}

{\bf Theorem 3.1}\hspace*{0.2cm}The following Lax pair
${\cite{Metin}}$,
\begin{eqnarray}
 && \hat{L}\varphi=(\partial+u\partial^{-1})\varphi=\lambda\varphi,
\label{KdVhchNL} \\
 &&\varphi_t=\hat{P}_{m}\varphi=(\hat{L}^{m})_{+}\varphi \label{KdVhchNP} 
\end{eqnarray}
generates the KdV hierarchy, where $m$ is an odd integer 
and  $\hat{L}^{m}_+$ is the differential part \cite{Dickey}  of 
pseudo-differential operator $\hat{L}^{m}$.

{\bf Proof:} First we  prove
\begin{eqnarray}
\hat{L}^{n}_{+}=\partial h,\label{LpDh}
\end{eqnarray}
where $n$ is an odd positive integer and
$h$ is a differential operator.
In fact for odd $n$ we will prove
\begin{eqnarray}
(\partial+u\partial^{-1})^{n}_{+}\partial=\partial
(\partial+\partial^{-1}u)^{n}_{+}. \label{DpDDDp}
\end{eqnarray}
By
$ \partial^{-1}(\partial+u\partial^{-1})^{n}\partial=
(\partial+\partial^{-1}u)^{n} $,
we obtain
$ (\partial+u\partial^{-1})^{n}\partial=\partial
(\partial+\partial^{-1}u)^{n} $.
Then we immediately get
 \begin{eqnarray}
[(\partial+u\partial^{-1})^{n}\partial]_{+}=[\partial
(\partial+\partial^{-1}u)^{n}]_{+}. \label{KdVposEQ}
 \end{eqnarray}
Equation (\ref{KdVposEQ}) is equivalent to
\begin{eqnarray}
(\partial+u\partial^{-1})^{n}_{+}\partial+{\rm
res}(\partial+u\partial^{-1})^{n}=
\partial(\partial+\partial^{-1}u)^{n}_{+}+{\rm
res}(\partial+\partial^{-1}u)^{n}. \label{eqv01}
\end{eqnarray}
In fact for odd $n$ we have
\begin{eqnarray}
{\rm res}\left( (\partial+u\partial^{-1})^{n} \right) ={\rm
res}\left( (\partial+\partial^{-1}u)^{n} \right) , \label{ResRes}
\end{eqnarray}
because
\begin{eqnarray}
{\rm res}\left( (\partial+u\partial^{-1})^{n} \right) ={\rm
res}\left( (-1)^{n}[(\partial+\partial^{-1}u)^{n}]^{*}\right)
=(-1)^{n+1}{\rm
res}\left( (\partial+u\partial^{-1})^{n} \right) . \nonumber
\end{eqnarray}
Formula (\ref{DpDDDp}) is just a direct result of 
Equation (\ref{eqv01}) and  Equation (\ref{ResRes}).

Secondly we prove that $\frac{d}{dt}\hat L=[\hat{P}_{m}, \hat{L}]$ is  
only one PDE for $u$, or in other words we will prove
\begin{eqnarray}
[\hat{P}_{m}, \hat{L}]=f[u] \partial^{-1}, \label{LPfD}
\end{eqnarray}  where $f[u]$ denotes
a differential polynomial of $u$.
By (\ref{LpDh}) 
\begin{eqnarray}
[\hat{P}_{m},\hat{L}]=\hat G+f[u] \partial^{-1}, \label{PLGf}
\end{eqnarray}
where $\hat G$ is a differential operator.
But we also have \cite{Dickey}
\begin{eqnarray}
   [\hat{P}_{m},\hat{L}]=[\hat{L}^{m}-\hat{L}^{m}_{-},\hat{L}]
   =-[\hat{L}^{m}_{-},\hat{L}]. \nonumber
\end{eqnarray}
So the order of $[\hat{P}_{m},\hat{L}]$ is
less than 0. This fact and Equation (\ref{PLGf}) imply Equation (\ref{LPfD}).
So we have proved that $\frac{d}{dt}\hat L=[\hat{P}_{m}, \hat{L}]$
is only a PDE $u_t=f[u]$.

At last we prove  $\frac{d}{dt}\hat L=[\hat{P}_{m}, \hat{L}]$,
$m=1,2,3,\cdots$,
is the KdV hierarchy. This can be verified by its recursion operator
 $\hat \varphi=\partial^2+4u+2u_{x}\partial^{-1}$,
which may be easily carried out by the method established by \cite{Metin}.

By Theorem 3.1 and Theorem 2.2 we immediately know
\begin{eqnarray}
\sum_{j=1}^m b_j (\partial+u\partial^{-1})^{j-1} u=0 \nonumber
\end{eqnarray}
is a proper constraint, which has been proved \cite{LLL} to be all the soliton
solutions of the KdV equation.

\section{Special constraints for the  AKNS hierarchy}

\subsection{The special type of Lax pair for the AKNS hierarchy}
The AKNS hierarchy \cite{AKNS, Li}
is
\begin{eqnarray}
\left(
  \begin{array}{c}  q\\ r \\ \end{array}
\right)_{t}
=\hat{\varphi}^{n}\left( \begin{array}{c} -iq \\ ir\\ \end{array} \right),
\nonumber
\end{eqnarray}
where $\hat{\varphi}$ is the recursion operator
 \begin{eqnarray}
\hat{\varphi}=\frac{1}{i}\left(
  \begin{array}{cc}
  -\partial+2q\partial^{-1}r & 2q\partial^{-1}q \\
 -2r\partial^{-1}r & \partial-2r\partial^{-1}q \\
  \end{array}
 \right) \nonumber
\end{eqnarray}
and $n$ is an arbitrary positive integer.  The first equation of
the hierarchy is
\begin{eqnarray}
   \left( \begin{array}{c} q \\ r \\ \end{array} \right)_{t}
=\frac{1}{i}\left( \begin{array}{c} -q_{xx}+2q^2r \\ r_{xx}-2qr^2 \\ \end{array} \right). 
\label{AKNSno1}
\end{eqnarray}

The natural Lax pair
\begin{eqnarray}
        \hat{\varphi}_t=[\hat P_m,  \hat \varphi] \nonumber
     \end{eqnarray}
of the AKNS hierarchy  fulfills  Theorem 2.2, 
where $\hat P_m$ is the linearization operator of the AKNS.
So the constraint $ \displaystyle
\sum_{j=1}^m b_j \varphi^{j-1} \left(\begin{array}{c}q\\-r
\end{array}\right)=
\left(\begin{array}{c}0\\0\end{array}\right) $  is a proper constraint.
But  here we will not discuss this useful symmetry constraint. 
We will investigate the constraints induced by the following Lax pair
of the AKNS.

{\bf Theorem 4.1} \cite{OS}\hspace*{0.2cm}
Let $\hat{L}=\frac{1}{i}(-\partial+q\partial^{-1}r)$,
$\hat{P}_{m}=\frac{1}{i}\hat{L}_{+}^{m}$. Then Lax equation
\begin{eqnarray}
\hat L_t=[\hat P_m, \hat L] \label{AKNSLax}
\end{eqnarray}
generates the AKNS hierarchy.

{\bf Proof}: First we must prove the Lax equation (\ref{AKNSLax})
is just two PDEs for $q$ and $r$ respectively.
In fact we will prove
\begin{eqnarray}
\left( \begin{array}{c} q \\ r\\ \end{array} \right)_{t}
=\left( \begin{array}{c} P_{m}\cdot q \\ -P_{m}^{*}\cdot r  \\ \end{array} \right),
\label{qrqrPqpr}
\end{eqnarray}
where $\hat P_{m}\cdot q$ denotes the differential polynomial
gotten by acting the operator $\hat P_m$ on $q$ and
$\hat P^*$ is the conjugate operator of $\hat P$.
Because
\begin{eqnarray}
[\hat{P}_{m},\hat{L}]=\frac{1}{i}[\hat{L}^{m}-\hat{L}^{m}_{-},\hat{L}]=
i[\hat{L}^{m}_{-},\hat{L}], \nonumber
\end{eqnarray}
we know $[\hat{P}_{m},\hat{L}]_+=0$.
So
\begin{eqnarray}
[\hat{P}_{m},\hat{L}]=(P_{m}\cdot
q)\partial^{-1}r-q\partial^{-1}(P_{m}^{*}\cdot r).
\label{CPLPqr}
\end{eqnarray}
With Equation (\ref{CPLPqr}) and  Equation (\ref{AKNSLax})
we immediately get Equation (\ref{qrqrPqpr}).

Secondly we will prove the recursion operator of the hierarchy (\ref{CPLPqr}) 
is 
\begin{eqnarray}
\hat \varphi=\frac{1}{i}\left( 
 \begin{array}{cc} -\partial+2q\partial^{-1}r & 2q\partial^{-1}q\\
            -2r\partial^{-1}r & \partial-2r\partial^{-1}q \\ \end{array}
   \right).  \label{varphiParqrqrrr}
\end{eqnarray}
With (\ref{varphiParqrqrrr}) Theorem 4.1 becomes obvious, while
Equation  (\ref{varphiParqrqrrr}) follows from the following
theorem.

{\bf Theorem 4.2}\hspace*{0.2cm} For any $n$
\begin{equation}
\hat{L}_{t_{n+1}}=\hat{L}_{t_{n}}\hat{L}+[\tilde R_n,\hat{L}],
\label{Ltn1LtnComm}
\end{equation}
where $\tilde R_n=\frac{1}{i}(a_n+b_n\partial^{-1}r)$.
Equation (\ref{Ltn1LtnComm}) is just the  recursion equation
\begin{eqnarray}
\left( \begin{array}{c}q_{t_{n+1}}\\r_{t_{n+1}} \end{array}\right) =
\hat \varphi \left( \begin{array}{c}q_{t_n}\cr r_{t_n} \end{array} \right),
\label{AKNSrecursionExpr}
\end{eqnarray}
where $\hat \varphi$ is the recursion operator (\ref{varphiParqrqrrr}).

{\bf Proof}: We will first prove Equation (\ref{Ltn1LtnComm}) .
Since $\hat{L}^{n+1}=\hat{L}\hat{L}^{n}$,  we have
\begin{eqnarray}
\hat{P}_{n+1}
=\frac{1}{i}(\hat{L}^{n+1})_{+}=\frac{1}{i}(\hat{L}^{n}\hat{L})_{+}=\frac{1}{i}
\left( ((\hat{L}^{n})_{+}\hat{L}+(\hat{L}^{n})_{-}\hat{L})_{+} \right)
=\frac{1}{i}\left( \hat{L}^{n}_{+}\hat{L}
-(\hat{L}^{n}_{+}\hat{L})_{-}+(\hat{L}^{n}_{-}\hat{L})_{+} \right) , 
\nonumber
\end{eqnarray} 
which leads  directly to
\begin{eqnarray}
\hat{L}_{t_{n+1}}
= [ \hat{P}_{n+1},\hat{L}]=\hat{L}_{t_{n}}\hat{L}+[\tilde{R_{n}},\hat{L}],
\nonumber
\end{eqnarray}
where $\tilde R_n$ has the expression
\begin{eqnarray}
\tilde{R_{n}}
=\frac{1}{i}\left(-(\hat{L}^{n}_{+}\hat{L})_{-}+(\hat{L}^{n}_{-}\hat{L})_{+}
\right). \nonumber
\end{eqnarray}
So $\tilde R_n$ can be  expressed as
\begin{eqnarray}
\tilde R_n=\frac{1}{i}(a_{n}+b_{n}\partial^{-1}r). \label{Rexpression}
\end{eqnarray}
Then we will prove (\ref{AKNSrecursionExpr}). 
Substituting (\ref{Rexpression}) to  (\ref{Ltn1LtnComm}) we get
\begin{eqnarray}
i \times \left( q_{t_{n+1}} \partial^{-1}r+ q \partial^{-1}r_{t_{n+1}} \right)
=(q_{t_n}\partial^{-1}r+q \partial^{-1} r_{t_n})
(-\partial+q \partial^{-1}r)
+ [a_n+b_n \partial^{-1}r, -\partial+q \partial^{-1}r]. 
\label{qtn1rtn1BBcomm}
\end{eqnarray}
The positive part of Equation (\ref{qtn1rtn1BBcomm}) gives
\begin{eqnarray}
a_n=\int (q_{t_n} r+q r_{t_n}) dx. \label{AKNSanEXPR}
\end{eqnarray}
Rearranging the negative part of Equation (\ref{qtn1rtn1BBcomm})
we get
\begin{eqnarray}
&&(i q_{t_{n+1}}-a_n q-b'_n) \partial^{-1}r
+q \partial^{-1} (i r_{t_{n+1}}-r'_{t_n}+r a_n) \nonumber\\
&=&(q_{t_n} +b_n) \partial^{-1} (r'+r q \partial^{-1} r)
+q \partial^{-1}(r_{t_n} q-r b_n)\partial^{-1}r.
\label{AKNSnegPRE}
\end{eqnarray}
Left multiplying (\ref{AKNSnegPRE})
with $\frac{1}{r} \partial$ and
right multiplying (\ref{AKNSnegPRE})
with $\partial \frac{1}{q}$ simultaneously,
considering its negative part we get 
\begin{eqnarray}
b_n=-q_{t_n}. \label{AKNSbnEXPR}
\end{eqnarray}
Substituting $b_n=-q_{t_n}$ to (\ref{AKNSnegPRE}) 
and considering $q_{t_n} r+q r_{t_n} =a'_n$, we  get
\begin{eqnarray}
(i q_{t_{n+1}}-2 a_n q-b'_n) \partial^{-1}r
+q \partial^{-1} (i r_{t_{n+1}}-r'_{t_n}+2 r a_n)
=0. \nonumber
\end{eqnarray}
Note we have applied the well-known formula 
$\partial^{-1}f'\partial^{-1}=f\partial^{-1}-\partial^{-1}f$.
Now it is clear
\begin{eqnarray}
q_{t_{n+1}}=\frac{1}{i} (2 a_n q+b'_n), \nonumber\\
r_{t_{n+1}}=\frac{1}{i} (r'_{t_n}-2 r a_n). \nonumber
\end{eqnarray}
By substituting (\ref{AKNSanEXPR}) and (\ref{AKNSbnEXPR}) to
the above equations we immediately get 
the recursion equation (\ref{AKNSrecursionExpr}).

\subsection{Specially  constraints for  AKNS hierarchy}
\hspace*{0.6cm}
Applying Theorem 2.2 to the Lax pair in Theorem 4.1 does not generate two systems of ODEs,
because the constraint $\hat L^n(0)=0$ only offers one constraint between $q$ and $r$
and still  another  constraint  between $q$ and $r$ must be given  for $q$ and $r$ being ODEs 
of the independent variable $x$. 
The situation is best explained by the case $n=1$. When $n=1$,by $\hat L (0)=0$ we immediately get
 $q=0$. So for the usual AKNS (\ref{AKNSno1}) only one PDE is left
$ r_{t}=-ir_{xx}$.

The second constraint   $\hat{L}^{2}(0)=0$ can be simplified to
\begin{eqnarray}
   qr=(\frac{q_{x}}{q})_{x}. \label{AKNScons2}
\end{eqnarray}
It has been noticed  \cite{Cheng} that 
with the  constraint (\ref{AKNScons2})
the  AKNS hierarchy is constrained to the Burgers hierarchy of $w$ 
by the transformation $w_{x}=qr=(\frac{q_{x}}{q})_{x}$.  And it is also
well-known that the Burgers
hierarchy can be linearized  by the famous Cole-Hopf transformation.
So with the  constraint (\ref{AKNScons2}) the AKNS hierarchy 
may be  linearized, which is best explained by the following theorem:

{\bf Theorem 4.3} With the  constraint (\ref{AKNScons2}) 
the AKNS hierarchy can be constrained to linear
equations $u_{t}=(-i)^{m-1}u^{(m)}$, where $u=\frac{1}{q}$.

{\bf Proof}: Let's prove this theorem by mathematical induction. When $k=2$
the AKNS, which is just the NLS equation in this case, is constrained to
\begin{eqnarray}
   iq_{t}+q_{xx}-2q(\frac{q_{x}}{q})_{x}=0.\nonumber
\end{eqnarray}
Substituting  $q$ with $\frac{1}{u}$, we get the linear equation
$u_{t_{2}}=-i u_{xx}$. When $k=n$, we assume the equation has been
transformed to $u_{t_{n}}=(-i)^{n-1}u^{(n)}$, i.e., 
\begin{eqnarray}
    \left(
\begin{array}{c} q_{t_{n}} \\ r_{t_{n}}\\ \end{array} \right)
=\hat{\varphi}^{n}\left(
 \begin{array}{c} -iq \\ ir \\ \end{array} \right) 
=\left( \begin{array}{c} 
   -(-i)^{n-1}q^{2}(\frac{1}{q})^{(n)} \\ 
\left( (\ln q)_{xx}/q\right)_{t_n}
    \end{array} \right). \nonumber
\end{eqnarray}
  When $k=n+1$, by $qr=(\frac{q_{x}}{q})_{x}$  we get
$\partial^{-1}(qr)_{t_{n}}=(\frac{q_{x}}{q})_{t_{n}}$, then by the
recursion operator we get
$$q_{t_{n+1}}=\frac{1}{i}\left(
   \begin{array}{cc}
  -\partial+2q\partial^{-1}r & 2q\partial^{-1}q \\ \end{array} \right)
  \hat{\varphi}^n \left(
  \begin{array}{c} -iq \\ ir \\ \end{array} \right).
$$
 Substituting  $q=\frac{1}{u}$, we finally we get $u_{t_{n+1}}=(-i)^{n}u^{(n+1)}$. 
This finishes the  proof.

Then we consider the constraint $\hat{L}^{3}(0)=0$, which is equivalent to
\begin{eqnarray}
   qr=(\frac{(q^3r)_{x}-qq_{xxx}+q_{x}q_{xx}}{-qq_{xx}+q_{x}^2+q^3r})_{x}
=\left(\ln(q^3r-qq_{xx}+q_{x}^2) \right)_{xx}. \label{AKNSconsL3}
\end{eqnarray}

{\bf Theorem 4.4}\hspace{0.2cm}With (\ref{AKNSconsL3})  the  standard AKNS 
 is constrained  to  $g_t=-i g_{xx}$ by the
transformation
\begin{eqnarray}
g=\frac{q}{q^3r-qq_{xx}+q_{x}^2}. \label{EXPRg}
\end{eqnarray}

{\bf Proof}:\hspace*{0.2cm} Direct calculation shows
$g_t+i g_{xx}=2 i g \left(qr-\left(\ln(q^3r-qq_{xx}+q_{x}^2) \right)_{xx}\right)$.

By (\ref{EXPRg}) and (\ref{AKNSconsL3}) $q$ can be easily obtained
\begin{eqnarray}
q=-\frac{1}{g (\ln g)_{xx}} .\label{qBYg}
\end{eqnarray}

Let us explain how (\ref{EXPRg}) is obtained.
Equation $iq_{t}+q_{xx}-2q^2r=0$ is transformed to
\begin{eqnarray}
&&-4u_{x}^2u_{t}^2+7u_{t}^2uu_{xx}-2u_{t}u^2u_{xxt}+2u_{xt}^2u^2-2u_{xxx}^2u^2
    +11iu_{t}uu_{xx}^2-2iu_{t}u^2u_{xxxx}-iu_{t}^3u   \nonumber\\
    &&-5u_{xx}^3u+4iu^2u_{xt}u_{xxx}-2iu^2u_{xxt}u_{xx}
    -8iu_{x}^2u_{t}u_{xx}+4u_{x}^2u_{xx}^2+2u_{xxxx}u^2u_{xx}=0 \label{AKNSc3EQu}
\end{eqnarray} 
by (\ref{AKNSconsL3}) and $q=\frac{1}{u}$. 
Here $u_{t}=-iu_{xx}$ is still a solutions of
 Equation (\ref{AKNSc3EQu}). 
Then Equation (\ref{AKNSc3EQu}) is transformed to
\begin{eqnarray}
 iv_{xxt}-iv_{xx}v_{t}-v_{xxxx}+2v_{xxx}v_{x}-v_{xx}v_{x}^2+3v_{xx}^2=0
\label{AKNSc3EQv}
\end{eqnarray}
by  $e^{v}=\frac{u_{t}+iu_{xx}}{u^2}$.
Now the solution $u_{t}=-iu_{xx}$ has been ruled out.
Equation (\ref{AKNSc3EQv}) can be written into a more  compact form
\begin{eqnarray}
i (\partial^2- v_{xx}) (v_{t}+i v_{xx}-i v_{x}^2)=0. \label{OPdecomV}
\end{eqnarray}

Equation (\ref{OPdecomV}) suggests us to calculate $v_{t}+i v_{xx}-i v_{x}^2$, 
which turned out to be $0$. But  $v_{t}+i v_{xx}-i v_{x}^2=0$ is linearized by
transformation $v=-ln(g)$. Therefore we get the final  transformation (\ref{EXPRg}).
Note that  $v_{t}+i v_{xx}-i v_{x}^2=0$  can  also be  gotten 
by considering the compatibility condition between equation $ir_{t}-r_{xx}+2qr^2=0$, 
 Equation (\ref{AKNSconsL3}) and (\ref{AKNSc3EQv}).

The third  equation of the  AKNS  hierarchy is the coupled KdV
\begin{eqnarray}
 \left(
\begin{array}{c}
q\\r
\end{array}
\right)_t=
\left( \begin{array}{c}
6 q r q_x-q_{xxx}\\6 q r r_x-r_{xxx}
\end{array}
\right). \label{COUPLkdv}
\end{eqnarray}

It is easy to verify that  under constraint (\ref{AKNSconsL3})  
Equation (\ref{COUPLkdv}) is simplified to $g_t=-g_{xxx}$, 
where $g$ is also defined by (\ref{EXPRg}).

Then we consider the case $\hat{L}^{4}(0)=0$, which is equivalent to
\begin{eqnarray}
q r=(\ln p_3)_{xx}, \label{AKNScons4}
\end{eqnarray}
where $p_3$ is defined as
\begin{eqnarray}
&&p_3=5q_{x}^2r^2q^3-q_{xxxx}qq_{xx}+q_{xxxx}q^3r-2q_{xxx}q^3r_{x}-6q_{xxx}q_{x}q^2r-2q_{xxx}q_{x}q_{xx} \nonumber \\
&&\hspace{0.8cm}+7q_{xx}q_{x}^2qr+8q_{x}r_{x}q^2q_{xx}+q_{xxxx}q_{x}^2+3q_{xx}^2q^2r-5r^2q_{xx}q^4-6q_{x}^3qr_{x}-q^5r_{xx}r
\nonumber \\
&&\hspace{0.8cm}+q^3r_{xx}q_{xx}-q^2q_{x}^2r_{xx}-5q_{x}^4r+q_{xxx}^2q+q^5r_{x}^2+q^6r^3+q_{xx}^3.
\label{p3DEF}
\end{eqnarray}
Let us define $p_2$ as
\begin{eqnarray}
p_2=q^3r-qq_{xx}+q_{x}^2. \label{p2DEF}
\end{eqnarray}
It can be direct verified that  under constraint (\ref{AKNScons4}) the standard AKNS is simplified to
$h=-i h_{xx}$, where $h=\frac{p2}{p3}$.
Then $q$ can be obtained  by
\begin{eqnarray}
\frac{1}{q}=h (\ln h)_{xx} \left( \ln(h \; h (\ln h)_{xx} )\right)_{xx}. \label{qASh}
\end{eqnarray}

Before summarizing all results above, let us first define
$p_k$ and $\theta_k(g)$  recursively.

Function $p_k$ is completely determined by $k$:
\begin{eqnarray}
&&p_{-1}=r, \quad p_0=1, \quad p_1=q,\nonumber\\
&&\frac{p_{k+1}}{p_k}=\left( q r-\left( \ln p_k \right)_{xx}\right) \frac{p_k}{p_{k-1}}.
\label{pkDEF}
\end{eqnarray}

Function $\theta_k(g)$ is  determined by  both $k$ and $g$:
\begin{eqnarray}
&&\theta_1(g)=g, \nonumber\\
&&\theta_{k+1}(g)= 
\left( \ln (\theta_1(g) \theta_2(g) \cdots \theta_k(g)) \right)_{xx} \theta_k(g).
\label{thetaDEF}
\end{eqnarray}

Now  we summarize  our main  result as following:

{\bf Theorem 4.5}\hspace{0.2cm}
The constraint $\hat L^{n}q=0$ is equivalent to $q r=(\ln p_n)_{xx}$,
where $\hat{L}=\frac{1}{i}(-\partial+q\partial^{-1}r)$.
With the constraint the $m$-th AKNS equation is equivalent to
$\frac{\partial u_n}{\partial t}=(-i)^{m-1} u_n^{(m)}$,
where $u_n=\frac{p_{n-1}}{p_n}$ and $p$ is defined by (\ref{pkDEF}).
Given  $u_n$, $q$ and $r$ are determined by
$q^{-1}=(-1)^{n-1}\theta_n (u_n) $ and
$r=(-1)^n\theta_{n+1} (u_n) $ respectively,
where $\theta$ is defined by (\ref{thetaDEF}).

The proof of  Theorem 4.5 consists of several parts.
The following theorems greatly reduce the complexity
of the proof of Theorem 4.5. 
So we will first prove the following
Lemma 4.6, Theorem 4.7 and Proposition 4.8, Theorem 4.9.
At last we will prove Theorem 4.5.

Let us define $(i \hat L)^k q=f_k$, and define $\alpha_j$ by 
$\alpha_0=q$ for $j=0$ and   $\alpha_j=q r- (\ln p_j)_{xx}$ for $j \neq 0$.

{\bf Lemma 4.6}\hspace{0.2cm}
$f_k=\alpha_0  \partial^{-1} \alpha_1 \partial^{-1} \alpha_2
\cdots \alpha_{k-1} \partial^{-1} \alpha_k$.

{\bf Proof:} Obviously the theorem is true for $k=1$.
Suppose the  theorem is true for $k=s$.
Then we must compute 
$f_{s+1}=(-\partial+q \partial^{-1} r) f_s$.
Let us  define $\hat C_j$ by
$\hat C_j=\alpha_j \partial^{-1}$.
Then we have $f_k=\hat C_0 \hat C_1 \cdots \hat C_k 0$, where
we have set $\partial^{-1} 0=1$.
It can be verified $\hat B_k \hat C_k=\hat C_k B_{k+1}$,
where $\hat B_k$  is defined as $\hat B_0=i \hat L$ for $k=0$
and $\hat B_k=-\partial -(\ln  
\frac{p_k}{p_{k-1}})_x+ 
\alpha_k \partial^{-1} $ for $k \neq 0$. 
Therefore, the theorem must be true for $k=s+1$, 
because $\hat B_{s+1} 0=\alpha_{s+1}$.

By Lemma 4.6, we immediately get:

{\bf Theorem 4.7}\hspace{0.2cm}
$\hat L^n q=0$ implies $q r=(\ln p_n)_{xx}$.

The following  proposition is also crucial for the proof of our final result.

{\bf Proposition 4.8}\hspace{0.2cm} For the AKNS   hierarchy
\begin{eqnarray}
\frac{(p_{k})_{ t_{s+1}}}{p_{k}}-\frac{(p_{k-1})_{t_{s+1} }}{p_{k-1}}
=\frac{1}{i} \left( 
2 \partial^{-1}(q r)_{t_s}-
\frac{(p_{k-1})_{t_s x}}{p_{k-1}}+
\frac{(p_{k})_x}{p_{k}} \frac{(p_{k-1})_{t_s}}{p_{k-1}}-
\frac{(p_{k})_{t_s x} }{ p_{k}}+
\frac{(p_{k-1})_x}{p_{k-1}} \frac{(p_{k})_{t_s}}{p_{k}}
\right)
\label{pt1B}
\end{eqnarray}
and
\begin{eqnarray}
(q r)_{t_{s+1}}-(\ln p_k)_{t_{s+1}xx}&=&
\frac{p_{k+1} p_{k-1}}{i \; p_k^2} \left(
\frac{(p_{k-1})_{t_s x}}{p_{k-1}}
-\frac{ (p_{k+1})_{t_s x} }{p_{k+1}}
+\frac{(p_{k+1})_x (p_k)_{t_s}}{p_{k+1} p_k}
-\frac{(p_k)_x (p_{k-1})_{t_s}}{p_k p_{k-1}} \right. \nonumber\\
&& \left. +\frac{(p_k)_x (p_{k+1})_{t_s}}{p_k p_{k+1}}
-\frac{(p_{k-1})_x (p_k)_{t_s}}{ p_{k-1} p_k}
\right) .
\label{pxxt}
\end{eqnarray}
{\bf Sketch of the proof:} According to the  recursion equation (\ref{AKNSrecursionExpr})
the proposition is clearly true for $k=1$.
So let us suppose the proposition is true for for $k=j$.
Then we must prove it is also true for $k=j+1$.
First we can solve $(p_{j-1})_{t_{n+1}}$  and  $(p_j)_{t_n x x} $
by (\ref{pt1B}) and (\ref{pxxt}) when $k=j$.
When $k=j+1$, we can verify, by a lengthy but straightforward calculations,
Equations (\ref{pt1B}) and (\ref{pxxt}) 
are simply  two identities  after considering (\ref{pkDEF}).

{\bf Theorem 4.9}\hspace{0.2cm}
If $q$ and $\theta$ are defined by (\ref{pkDEF}) and 
(\ref{thetaDEF}) respectively. 
Given arbitrary function $g$, set $\theta_1=g$,
$q^{-1}=(-1)^{n-1}\theta_n (g) $ and
$r=(-1)^n\theta_{n+1} (g) $.
Then $\theta_{n-k+1}=(-1)^{n-k} \frac{p_{k-1}}{p_k}$.
As a result, $g=\frac{p_{n-1}}{p_n}$, $q r=\left( \ln p_n \right)_{xx}$.
Conversely if $p$ is defined by (\ref{pkDEF}),
$\theta$ is defined by $\theta_{n-k+1}=(-1)^{n-k} \frac{p_{k-1}}{p_k}$
and  there is a constraint
$q r=\left( \ln p_n \right)_{xx}$ between $q$ and $r$,
then  $\theta$ satisfies (\ref{thetaDEF}), 
$q^{-1}=(-1)^{n-1}\theta_n (g) $ and  $r=(-1)^n\theta_{n+1} (g) $,
where $g=\frac{p_{n-1}}{p_n}$.

{\bf Proof}: The proof  is obvious by mathematical induction.

Now we will give the proof of Theorem 4.5.

{\bf  Proof of Theorem 4.5:} We only need to  prove
\begin{eqnarray}
\frac{\partial u_n}{\partial t}=(-i)^{m-1} u_n^{(m)}. \label{unt}
\end{eqnarray}
In fact (\ref{unt}) is a direct result of the following equation
\begin{eqnarray}
(u_n)_{t_{k+1}}=-i \times (u_n)_{t_{k} x},
\label{utn}
\end{eqnarray}
because (\ref{unt}) is obviously true for $m=1$.
Suppose (\ref{utn}) has  been true for $m=k$.

Then $u_{t_{k+1}}$ can be directly calculated as
\begin{eqnarray}
u_{t_{k+1}}&=&-\frac{p_{n-1}}{p_n}\left( 
\frac{(p_n)_{t_{k+1}}}{p_n}-\frac{(p_{n-1})_{t_{k+1}}}{p_{n-1}}\right) \nonumber\\
&=&i \frac{p_{n-1}}{p_n}
\left( 
2 \partial^{-1}(q r)_{t_k}-
\frac{(p_{n-1})_{t_k x}}{p_{n-1}}+
\frac{(p_{n})_x}{p_{n}} \frac{(p_{n-1})_{t_k}}{p_{n-1}}-
\frac{(p_{n})_{t_k x} }{ p_{n}}+
\frac{(p_{n-1})_x}{p_{n-1}} \frac{(p_{n})_{t_k}}{p_{n}}
\right)\nonumber\\
&=&
i \frac{p_{n-1}}{p_n}
\left( 
2 (\ln p_n)_{{t_k}x}-
\frac{(p_{n-1})_{t_k x}}{p_{n-1}}+
\frac{(p_{n})_x}{p_{n}} \frac{(p_{n-1})_{t_k}}{p_{n-1}}-
\frac{(p_{n})_{t_k x} }{ p_{n}}+
\frac{(p_{n-1})_x}{p_{n-1}} \frac{(p_{n})_{t_k}}{p_{n}}
\right)\nonumber\\
&=&-i\left(-\frac{p_{n-1} (p_n)_{t_k}}{p_n^2} +
\frac{(p_{n-1})_{t_k}}{p_n} \right)_x \nonumber\\
&=&-i \; u_{t_k x} . \nonumber
\end{eqnarray}
So Equation (\ref{utn}) is proved.
%But Equation (\ref{utn}) is a direct result
%of equation $q r=(\ln p_n)_{xx}$ and  Equation (\ref{pt1B}).

Let us  give a simple example to illustrate Theorem 4.5.

Choose $n=4$ and $m=3$ in Theorem 4.5.
Then $\frac{u_4}{\partial t}=-(u_4)_{xxx}$.
Choose a solution of it such as
$u_4=1+e^{8 t-2 x}+e^{t-x}+e^{x-t}+e^{2 x-8 t} $.
Then we obtain
\begin{eqnarray}q&=&
\frac{-e^{18 t}-e^{6 x}-9 e^{17 t+x}-36 e^{10 t+2 x}-9 e^{16 t+2 x}-65 e^{9
    t+3 x}-9 e^{2 t+4 x}-36 e^{8 t+4 x}-9 e^{t+5 x}}{36 e^{17 t+x}+576 e^{10
    t+2 x}+1296 e^{9 t+3 x}+576 e^{8 t+4 x}+36 e^{t+5 x}}, \nonumber\\
r&=&\frac{576 e^{8 t+2 x}}{e^{16 t}+e^{4 x}+16 e^{9 t+x}+36 e^{8 t+2 x}+16 e^{7
    t+3 x}}. \nonumber
\end{eqnarray}
It is easy to  check that the above solution is indeed a solution of
(\ref{COUPLkdv}), the  $m$-th equation of AKNS hierarchy.

\section{Conclusions}
\hspace*{0.6cm}In this work, we apply a special form of Lax pair  to analyze the 
solutions of  KdV hierarchy and AKNS hierarchy. For the KdV hierarchy,  the soliton
solutions are completely depicted by a new, simple and direct way.
For the AKNS Hierarchy a special form of Lax pair is  analyzed.
With the special kind of Lax pair a wide class of solutions of the AKNS hierarchy have been obtained.
At last we  give the main theorem  which states  how to linearize all equations of the AKNS hierarchy
with the special kind of constraints mentioned in this paper.

\section*{Acknowledgments}
\hspace*{0.6cm}
The authors would like to thank Prof. Lou S. Y., who had read the draft and given a lot of
instructions.
The work is partly supported by  NSFC (No.10735030), NSF of Zhejiang Province (R609077,
Y6090592),
NSF of Ningbo City (2009B21003, 2010A610103, 2010A610095).

\end{document}